\documentclass[preprint,prd,superscriptaddress,aps, nofootinbib,tightenlines]{revtex4-2}

\usepackage{amsmath}
\usepackage{amssymb}
\usepackage{physics}
\usepackage{feyn}
\usepackage{enumerate}
\usepackage{graphicx}
\usepackage{xcolor}
\usepackage{dcolumn}
\usepackage{bm}
\usepackage{bbm}
\usepackage{hyperref}
\usepackage[mathlines]{lineno}
\usepackage{mathrsfs,subfigure}
\usepackage{booktabs}
\usepackage[normalem]{ulem}

\allowdisplaybreaks 

\hypersetup{colorlinks=true,citecolor=black, linkcolor=black, urlcolor = black,filecolor=black}

\begin{document}

\hfill{ {\small USTC-ICTS/PCFT-24-35} }

\title{Scalar-Gauss-Bonnet gravity: Infrared causality and detectability of GW observations}

\author{Wen-Kai Nie}
\email{wknie@mail.ustc.edu.cn}

\author{Lin-Tao Tan}
\email{lttan@ustc.edu.cn}

\affiliation{Interdisciplinary Center for Theoretical Study, University of Science and Technology of China, Hefei, Anhui 230026, China}
\affiliation{Peng Huanwu Center for Fundamental Theory, Hefei, Anhui 230026, China}

\author{Jun Zhang}
\email{zhangjun@ucas.ac.cn}

\affiliation{International Centre for Theoretical Physics Asia-Pacific, University of Chinese Academy of Sciences, Beijing 100190, China}
\affiliation{Taiji Laboratory for Gravitational Wave Universe (Beijing/Hangzhou), University of Chinese Academy of Sciences, Beijing 100049, China}

\author{Shuang-Yong Zhou}
\email{zhoushy@ustc.edu.cn}

\affiliation{Interdisciplinary Center for Theoretical Study, University of Science and Technology of China, Hefei, Anhui 230026, China}
\affiliation{Peng Huanwu Center for Fundamental Theory, Hefei, Anhui 230026, China}

  \newcommand{\nie}[1]{{\color{red}#1}}

\date{\today}

\begin{abstract}
We investigate time delays of wave scatterings around black hole backgrounds in scalar-tensor effective field theories of gravity. The scalar-Gauss-Bonnet (sGB) couplings, being corrections of the lowest orders, can give rise to hairy black holes. By requiring infrared causality, we impose lower bounds on the cutoff scales of the theories. With these bounds, we further discuss the detectability of sGB gravity in gravitational waves from binary black hole mergers. Compared with the gravitational effective field theories that contain only the two tensor modes, adding extra degrees of freedom, such as adding a scalar, opens up a detectable window in the planned observations.
\end{abstract}

\maketitle

\newpage

\tableofcontents

\section{Introduction and summary}
\label{intro}

The Scalar-Gauss-Bonnet (sGB) gravity has been extensively studied for its interesting phenomenological differences from general relativity (GR) in strong gravity regimes. In particular, the sGB couplings can induce hairy black holes (BHs)~\cite{Kanti:9511071dilBH,Torii:9606034DilatonBH,Yunes:1101.2921BHESGBdCS,Pani:1109.3996SlowlyrotatingBHEdGBdCS,Sotiriou:1312.3622sthair,Sotiriou:1408.1698sthair,Ayzenberg:1405.2133edgbSlowly-RotatingBH,Maselli:1507.00680RotatingBHEdGB} and give rise to spontaneous scalarization~\cite{Doneva:1711.01187Curvature-InducedScalarization, Silva:1711.02080ESGBSpontaneousscalarization, Antoniou:1711.03390ESGBNohair,Minamitsuji:1812.03551GBScalarizedBH,Silva:1812.05590StabilityScalarizedSGBBH,Macedo:1903.06784Self-interactionsScalarization}. The nature of such BHs, such as stability and quasi-normal modes, and gravitational waves (GWs) emissions from binary BHs in sGB gravity have also been studied~\cite{Mignemi:9212146edgbBH,Kanti:9703192EDGBBHStability,Zwiebach:1985EDGBString,Metsaev:1987dil,Ripley:1911.11027BHdynamicsESGB,Evstafyeva:2212.11359ringdownESGB,Pani:0902.1569EDGBBH,Julie:2202.01329esgbbhbinarysensitive,Yagi:1110.5950Inspirals,Blazquez-Salcedo:1609.01286edgbGWringdown,Pierini:2103.09870edgbqnm1storder,Bryant:2106.09657SGBpEq,Pierini:2207.11267edgbqnm2ndorder,Nojiri:2023mbo,Almeida:2024cqz,Antoniou:2024gdf,Antoniou:2024hlf}. With probes to the strong gravity regime offered by future observations, gravitational theories including sGB gravity can be tested with unprecedented precision~\cite{Mishra:1005.0304aLIGO&ET,Arun:0604018GWs,Yagi:1204.4524XRayesgb,Wang:2104.07590EDGBGW19,Nair:1905.00870gwtc1BBH,Yamada:1905.11859GWsTestingmassivefield,Tahura:1907.10059BBHGWAmplitudeCorrections,Perkins:2104.11189GWC,Wang:2302.10112ringdown,Gao:2405.13279edgbgw23,Lyu:2201.02543NSBHGWs,Saffer:2110.02997BNSEDGB,Liu:2023spp,Chen:2024ery}.

From the effective field theory (EFT) point of view, sGB gravity relies on some of the lowest order high-dimensional operators. While low-energy EFTs provide an efficient framework for searching for new physics in observations, their validity and consistency must be justified from theoretical considerations. For instance, it has long been known that many low-energy EFTs manifest superluminal propagation in curved spacetimes~\cite{Adams:0602178Causality,Shore:9504041Causalityanomaliesandhorizons,Shore:0012063photonswithgraity,Hollowood:0905.0771QEDCcurvedcausality, Hollowood:1512.04952CausalityUVCompletion,deRham:1909.00881Speed,deRham:2007.01847Causality,deRham:2005.13923GREFTBHGW,CarrilloGonzalez:2022fwg,CarrilloGonzalez:2023cbf,Serra:2023nrn,Goon:2016une,Benakli:2015qlh,Drummond:1979pp,Lafrance:1994in,Shore:2002gw,Hollowood:2007ku,Hollowood:2008kq,AccettulliHuber:2020oou,Edelstein:2021jyu,Benakli:2015qlh}. 
These superluminal propagations, however, are unresolvable within the EFT. In particular, one can impose the so-called asymptotic causality~\cite{Camanho:2014apa,Camanho:2016opx,Goon:2016une,Hinterbichler:2017qcl,Hinterbichler:2017qyt,AccettulliHuber:2020oou,Gao:2000ga,Serra:2022pzl,Bellazzini:2021shn}, which argues that the speed of all species in an EFT cannot be secularly superluminal for the theory to be causal. In the case of wave scattering, asymptotic causality requires that the net time delay caused by the scattering, if resolvable, must be positive. However, it has recently been pointed out in  Refs.~\cite{deRham:2007.01847Causality,Chen:2112.05031causality,deRham:2112.05054Causality,Chen:2309.04534Causalityshockwaves, CarrilloGonzalez:2023emp, Melville:2024zjq,Chen:2309.04534Causalityshockwaves,Hollowood:1512.04952CausalityUVCompletion,Drummond:1979pp,Hollowood:0707.2302Causality,Hollowood:1512.04952CausalityUVCompletion,Shore:9504041Causalityanomaliesandhorizons,Shore:0012063photonswithgraity,Hollowood:0905.0771QEDCcurvedcausality,deRham:1909.00881Speed} that asymptotic causality does not fully capture all causality conditions 
available within EFTs (see Ref.~\cite{deRham:2022hpx} for a review). Instead of the net time delay, one can require the EFT corrections on all resolvable time delays to be positive.
This criterion is called infrared causality and is based on the following reasoning: Causality requires any support outside the light cone determined by the background geometry to be unresolvable, and the causal structure of the background geometry can be seen by high-frequency modes, which are only sensitive to the local inertial frame. In the case of scattering, it is the EFT corrections on the time delay that reflect the differences between the low- and high-frequency modes. In particular, a negative EFT correction indicates support outside the light cone seen by the high-frequency modes, and hence should not be resolvable in a causal EFT. Infrared causality is so powerful that it indicates that the gravitational EFTs that only contain the two tensor modes cannot be tested with current GW observations~\cite{deRham:2112.05054Causality}. However, as we shall demonstrate, adding extra field degrees of freedom can render a gravitational EFT testable in the upcoming GW experiments, while satisfying the causality bounds.

It is worth noting that the stronger infrared causality is sometimes required to obtain bounds on gravitational EFT coefficients that match the strength of those derived from positivity bounds via dispersion relations. For example, in pure Einstein EFT, while the parametric equivalence between dispersive positivity and asymptotic causality can be established for the bound on the cubic Riemann coefficient \cite{Caron-Huot:2201.06602Causality}, dispersive positivity requires the quartic Riemann coefficient to be positive \cite{Caron-Huot:2201.06602Causality}, which is consistent with infrared causality \cite{deRham:2112.05054Causality}, while asymptotic causality allows this coefficient to be negative. As another example, it is shown in Ref.~\cite{Chen:2112.05031causality}, for the Goldstone mode operator $c (\nabla \phi)^4 /\Lambda^D$ in D-dimensional spacetime, infrared causality requires $c \gtrsim-\left(\Lambda/M_{\mathrm{Pl}}\right)^{(D-2)}$, which is consistent with gravitational positivity bounds from Refs.~\cite{Haring:2023zwu,Alberte:2020jsk}. In contrast, asymptotic causality yields $c \gtrsim-\left(\Lambda/M_{\mathrm{Pl}}\right)^{(D-2) /2}$, which is a weaker constraint given that $\Lambda/M_{\mathrm{Pl}}\ll 1$.

The validity of infrared causality in certain kinematical regimes has been questioned for loop-level UV completions in Ref.~\cite{Bellazzini:2021shn}, which considered the case of QED minimally coupled to gravity as the partial UV completion. However, as pointed out in Ref.~\cite{Bellazzini:2021shn}, those regimes do not encompass the standard astronomical scenarios with black holes, which, in contrast, are the focus of our current work. Indeed, the loop calculations in Ref.~\cite{Bellazzini:2021shn} use flat-space matter propagators, while neglecting the higher-order (classical) gravitational/post-Minkowskian corrections, which are important in astronomical contexts and whose inclusion could restore infrared causality.

sGB gravity has been tested with various astrophysical observations, such as low-mass X-ray binary orbital decay, binary compact object mergers, and neutron star measurements~\cite{Yagi:1204.4524XRayesgb,Wang:2104.07590EDGBGW19,Nair:1905.00870gwtc1BBH,Yamada:1905.11859GWsTestingmassivefield,Tahura:1907.10059BBHGWAmplitudeCorrections,Perkins:2104.11189GWC,Wang:2302.10112ringdown,Gao:2405.13279edgbgw23,Lyu:2201.02543NSBHGWs,Saffer:2110.02997BNSEDGB}. It has also been constrained by positivity/causality bounds based on the dispersion relations of Poincaré invariant scattering amplitudes that connect the EFT with (unspecified) UV completions that are unitary and causal~\cite{Hong:2304.01259PobstEFT,Xu:2024iao}, which gives rise to bounds that are mostly independent of the EFT cutoff (see, for example, Refs.~\cite{deRham:2017avq, deRham:2017zjm, Arkani-Hamed:2020blm, Paulos:2017fhb, Bellazzini:2020cot, Tolley:2020gtv, Caron-Huot:2020cmc, Sinha:2020win, Zhang:2020jyn, Remmen:2020vts, Guerrieri:2020bto, Alberte:2020jsk, Tokuda:2020mlf, Li:2021lpe, Bern:2021ppb, Alberte:2021dnj, Chiang:2021ziz, Caron-Huot:2021rmr, Caron-Huot:2201.06602Causality, Henriksson:2022oeu, Haring:2022sdp} for some recent developments along this direction and  Ref.~\cite{deRham:2022hpx} for a review). 

In this work, we investigate the infrared causality constraints on sGB gravity by considering the scattering of GWs and scalar waves on BHs. We derive the master equations for the linear metric and scalar field perturbations on a static and spherically symmetric BH background, and find that the even gravitational mode is coupled with the scalar mode. For the odd mode, we can apply the garden-variety WKB approximation to calculate the time delays, while for the even case a multi-variable WKB method is utilized. By imposing infrared causality, we constrain the EFT corrections via the time delays and impose lower bounds on the EFT cutoff of sGB gravity. These lower bounds on the EFT cutoff strongly constrain the parameter space tested in the current and upcoming GW experiments. We also discuss the effects of adding the cubic curvature operator and more scalar degrees of freedom, and find that the lower bounds on the EFT cutoff remain almost unchanged. 

Then, we consider the observability of sGB gravity, using the infrared causality constraints as theoretical priors. Compared to the pure gravitational EFT case, we find that a detectable window opens up when the theory is endowed with an extra scalar degree of freedom that can lead to hairy BHs. This is largely due to the fact that the hairy black holes can give rise to dipole radiations. While this window appears narrow in the 2D plot of cutoff-vs-BH mass, we emphasize that the range of testable BH masses is actually quite large, spanning several orders of magnitude. For example, for an equal-mass binary BH of total mass $M_{\mathrm{tot}}$ at 300 Mpc, there exists a detectable window with $M_{\mathrm{tot}} \in [ 10, 10^3 ] \,M_{_{\odot}}$ or $\Lambda \in [ 10^{-10}, 10^{-7} ] \,\mathrm{eV}$ for the LISA experiment, where $M_{_{\odot}}$ is the mass of the sun and $\Lambda$ is the EFT cutoff. Additionally, we find that when more scalars ({\it i.e.}, multi-sGB couplings) are added, the detectable window can be enlarged. In short, ensuring the gravitational EFT is consistent with causality implies that any future detection of beyond-Einstein effects in GW experiments would be a clear sign of additional nontrivial degrees of freedom in strong gravity regime.

The paper is organized as follows. In Sec.~\ref{sec:BHsGB}, we provide an overview of sGB gravity and analytically derive the static and spherically symmetric BH background solution, including its scalar profile, by solving the field equations perturbatively. In Sec.~\ref{sec:scatter}, to study GWs and scalar waves propagating on the BH background, we use the Regge-Wheeler formalism to obtain the master equations for the linear metric and scalar field perturbations. Based on these master equations, we apply the WKB approximation to determine the phase shifts of the scattered waves. Using these phase shifts, we calculate the corresponding time delays, particularly their leading-order beyond-Einstein corrections. In Sec.~\ref{sec:causal}, we impose infrared causality to constrain sGB couplings and, consequently, the cutoff of sGB gravity. We also consider the effects of including the cubic curvature operator or multi-sGB couplings. Finally, in Sec.~\ref{sec:detect}, we discuss the detectability of sGB gravity in the upcoming GW experiments by identifying a detectable window, using infrared causality as a theoretical prior. We also explore the detectability when multi-sGB couplings are present.

\section{BHs in sGB gravity}
\label{sec:BHsGB}

\subsection{Theory}

The action of sGB gravity (with the natural units $\hbar=c=1$) is 
\begin{align}\label{ESGBAction}
   S=\frac{M_{\mathrm{Pl}}^2}{2}  \int \mathrm{d}^4 x \sqrt{-g} \left[R-\frac{1}{2}\partial _\mu \phi\partial ^\mu \phi+\frac{1}{\Lambda^2} f(\phi)  \mathcal{G}\right]  ,
\end{align}
where $M_{\text{Pl}}$ is the Planck mass and $\phi$ is a real scalar field that couples to the Gauss-Bonnet invariant $\mathcal{G} \equiv R_{\mu \nu \alpha \beta}^2-4 R_{\mu \nu}^2+R^2$ through a dimensionless coupling function $f(\phi)$ suppressed by a cutoff scale $\Lambda$. Variation of the action~Eq.\eqref{ESGBAction} yields the field equations as
\begin{align}
   \label{ESGBEOM1}
   G_{\mu \nu} & = T_{\mu \nu}, \\
   \label{ESGBEOM2}
   \square \phi & =- \frac{1}{\Lambda^2} f^{\prime}(\phi) \mathcal{G},
\end{align}
where $G_{\mu \nu}=R_{\mu \nu}-\frac{1}{2} g_{\mu \nu} R$, and $T_{\mu \nu}$ is the stress-energy tensor given by 
\begin{equation}
T_{\mu \nu}=\frac{1}{2} \nabla_\mu \phi \nabla_\nu \phi-\frac{1}{4}g_{\mu \nu}(\nabla \phi)^2-\frac{4}{\Lambda^2} P_{\mu \alpha \nu \beta} \nabla^\alpha \nabla^\beta f
\end{equation}
where $\nabla_\mu$ denotes the covariant derivative and $\square \equiv \nabla^\mu \nabla_\mu$. Here, a prime denotes differentiation with respect to the scalar field, i.e., $f^{\prime}(\phi) \equiv d f(\phi) / d \phi$. The  tensor $P_{\mu \nu \rho \sigma}$ is defined as $P_{\mu \nu \rho \sigma}=R_{\mu \nu \rho \sigma}-2 g_{\mu[\rho} R_{\sigma] \nu}+2 g_{\nu[\rho} R_{\sigma] \mu}+g_{\mu[\rho} g_{\sigma] \nu} R$.

To keep the discussion general, we shall take the EFT perspective, and expand the coupling function as
\begin{align}
    f(\phi)=c_1 \phi + c_2 \phi^2 + \cdots  ,
\end{align}
where $c_1$ and $c_2$ are the dimensionless coupling constants, and the dots stand for higher-order terms that are negligible in the small-$\phi$ expansion. In this expansion, without loss of generality, we have chosen the asymptotic value of $\phi$ to be zero. We shall consider both the $c_1$- and $c_2$-term, which have been extensively studied out of phenomenological interests. For example, a non-trivial $c_1$-term always dresses BHs with scalar hair~\cite{Yunes:1101.2921BHESGBdCS,Yagi:1110.5950Inspirals,Sotiriou:1312.3622sthair,Sotiriou:1408.1698sthair}, and with a $c_2$-term alone (or generally, coupling with $f^\prime(\phi_0)= 0$ and $f^{\prime \prime}(\phi_0) R_{\mathrm{GB}}^2>0$ at certain $\phi_0$), spontaneous scalarization can be triggered for compact objects~\cite{Silva:1711.02080ESGBSpontaneousscalarization,Doneva:1711.01187Curvature-InducedScalarization,Minamitsuji:1812.03551GBScalarizedBH,Silva:1812.05590StabilityScalarizedSGBBH,Macedo:1903.06784Self-interactionsScalarization}. In the latter case, GR BHs are solutions to sGB gravity, but the BHs evolve to become hairy due to tachyonic instabilities. 

\subsection{The background BH spacetime}

In this work, we focus on static and spherically symmetric BHs in sGB gravity, the metric of which can be written as
\begin{align} 
   \mathrm{d} s^2=  A(r) \mathrm{d} t^2 + B^{-1}(r) \mathrm{d} r^2 +r^2 \mathrm{d} \Omega^2  ,
\end{align}
while the scalar hair, if present, is given by $\bar{\phi}(r)$. It is convenient to define a dimensionless parameter 
\begin{align}
   \alpha \equiv (G M \Lambda)^{-2}, 
\end{align}
with $M$ being the Arnowitt-Deser-Misner mass of the BH and $G=(8 \pi M_{\mathrm{Pl}}^2)^{-1}$ being Newton's constant. For the EFT to be valid on scales down to the BH horizon, it generally requires $\Lambda^{-1} \ll GM$, that is, $\alpha \ll 1$. In this case, the sGB couplings manifest themselves as perturbative corrections to the Einstein-Hilbert term, and the BH solution can be obtained by solving the field equations order by order in $\alpha$,
\begin{align}
  \label{background1}
  A(r) &= A_0 (r) + \alpha^2 c_1^2 A_1 (r) + \mathcal{O}(\alpha^3)  , \\ 
  \label{background2}
  B(r) &=  B_0 (r) + \alpha^2 c_1^2 B_1 (r) + \mathcal{O}(\alpha^3)  , \\
  \label{background3}
  \bar{\phi}(r)&=   \alpha c_1   \phi_1 (r)+ \alpha^2 c_1  c_2  \phi_2 (r) + \mathcal{O}(\alpha^3)  ,
\end{align} 
with $A_0(r)=B_0(r)=1-2GM/r$ being the components of the Schwarzschild metric. The explicit expressions of $A_1$, $B_1$, $\phi_1$, and $\phi_2$ can be found in Appendix~\ref{app:Bg&mastereqsdetails}.

\section{Wave scattering and time delay}
\label{sec:scatter}

\subsection{BH perturbations in sGB gravity}

GWs propagating on a BH background can be treated as metric perturbations and studied with BH perturbation theory. In GR, the metric perturbations can be expressed with spherical harmonics and decomposed into odd and even parity modes in the frequency domain. Modes of different degrees, parity, or frequency decouple at linear order and propagate independently on the BH background. In particular, the radial dependence and hence the dynamics of each mode can be captured by master variables $\Psi_{\omega \ell}^{\pm}(r)$ that satisfy the well-known Regge-Wheeler-Zerilli equations~\cite{Regge:1957RWEq,Zerilli:1970ZEq}. Here, $\omega$ is the frequency, $\ell$ is the degree of the spherical harmonics, and $-/+$ denotes the odd/even parity. Given the spherical symmetry of the background, there is no dependence on the order $m$ of the spherical harmonics.

In sGB gravity, there are also scalar waves, $\delta \phi = \phi - \bar{\phi}$. 
To discuss wave scattering, we perform the same decomposition for metric perturbations as in GR and also express the scalar waves with spherical harmonics in the frequency domain. Then we find that the odd modes, involving only metric perturbations, propagate independently on the background, while the modes of even metric perturbations and of the scalar field generally couple with each other because of the sGB couplings. By carefully choosing the master variables (see Appendix~\ref{app:Bg&mastereqsdetails}), the odd and even master equations can be written as 
\begin{align} 
    \label{mastereqodd} 
    \frac{\mathrm{d}^2 \Psi_{\omega \ell}^{-}}{\mathrm{d} r_*^2}+\left(\omega^2- V^{ -} \right) \Psi_{\omega \ell}^{-} &=0,
\end{align} 
and
\begin{align}
    \label{mastereqeven} 
    \frac{\mathrm{d}^2}{\mathrm{d}r_{*}^{2}}\begin{pmatrix} \tilde{\Psi}_{\omega \ell}^{+} \\ \addlinespace \tilde{\Phi}_{\omega \ell}  \\\end{pmatrix}
    + \big( \omega^{2} \mathbbm{1} - \tilde{\mathbb{V}}\big) \begin{pmatrix} \tilde{\Psi}_{ \omega \ell} ^{+} \\ \addlinespace \tilde{\Phi}_{\omega \ell}  \\\end{pmatrix} &=0,
\end{align}
where $\Psi_{\omega \ell}^{-}$ is the odd master variable, $\tilde{\Psi}_{\omega \ell}^{+}$ and $\tilde{\Phi}_{\omega \ell}$ are the even master variables, and $r_*$ is the tortoise coordinate defined by $\mathrm{d} r_{*}=\mathrm{d}r/ \sqrt{A(r) B(r)} $. Here, $V^{-} $ is the odd potential, $\mathbbm{1}$ is a $2 \times 2$ identity matrix, and $\tilde{\mathbb{V}}$ is the $2 \times 2$ even potential matrix. Up to $\mathcal{O}(\alpha^2)$, we have
\begin{align}
    \label{potodd}
     V^{-}&=V_{\mathrm{GR}}^{-}+\alpha^{2} c_1^{2} \delta V^{-}_{1}, 
\end{align}
and
\begin{align}
     \label{potmat}
    \tilde{\mathbb{V}}&= \begin{pmatrix} V_{\mathrm{GR}}^{+}  & 0 \\ \addlinespace 0  & V^{(0)}_{\mathrm{GR}}   \\\end{pmatrix}+ \alpha \tilde{\mathbb{V}}_{1} + \alpha^{2} \tilde{\mathbb{V}}_{2},
\end{align}
where $V_{\mathrm{GR}}^{\pm}$ is the Regge-Wheeler-Zerrili potentials in GR, $V^{(0)}_{\mathrm{GR}} $ is the potential of a minimally coupled scalar field propagating on a Schwarzschild background, $\delta V^{-}_{1}$ is the correction on the odd potential, and $\tilde{\mathbb{V}}_{1}$ and $\tilde{\mathbb{V}}_{2}$ are the correction matrices for the even master equations, which are non-diagonal when $c_{1} \neq 0$ (see Appendix~\ref{app:Bg&mastereqsdetails} for their explicit expressions). 

\subsection{Wave scattering and time delay}

With the odd and even master equations Eqs.~\eqref{mastereqodd} and~\eqref{mastereqeven}, we can discuss wave scattering on the BH background and compute the resulting phase shift and time delay. In the case of wave scattering, $\gamma \equiv \omega^2/V_{\mathrm{max}} < 1$, where $V_{\mathrm{max}}$ denotes the maximum of the corresponding GR potential, i.e., $V_{\rm GR}^{\pm}$ or $V^{(0)}_{\rm GR}$. We demand that all the modes decay exponentially as they approach the BH horizon (tortoise coordinate $r_*$ goes to $- \infty$). Consequently, the desired WKB solutions for Eqs.~\eqref{mastereqodd} and~\eqref{mastereqeven} are
\begin{align}
  \Psi_{\omega \ell}^{-} & \simeq   \frac{ C^{-}}{(V^{-}-\omega^{2})^{\frac{1}{4}}} e^{- I^{-} } \qquad   (r_{*} < r_{*\mathrm{T}}^{-}), 
  \end{align}
and  
\begin{align}
  \begin{pmatrix} \tilde{\Psi}_{\omega \ell}^{+} \\ \addlinespace \tilde{\Phi}_{\omega \ell}  \\\end{pmatrix} & \simeq   \sum_{(k)=+,(0)} \frac{ C^{(k)} \vec{v}_{(k)}}{(V^{(k)}-\omega^{2})^{\frac{1}{4}}} e^{- I^{(k)} } \quad (r_{*} < r_{*\mathrm{T}}^{+,(0)}),
\end{align}
where 
\begin{align}
    I^{\pm,(0)} = \int_{r_{*}}^{r_{*\mathrm{T}}^{\pm,(0)}} \sqrt{V^{\pm,(0)}-\omega^2} \, \mathrm{d} r_* .
\end{align}
Here, $C^{\pm,(0)}$ are the integration constants, and  $V^-$ is given by Eq.~\eqref{potodd}. The terms $\omega^2-V^{+,(0)}$ represent the eigenvalues of the matrix $\omega^2-\tilde{\mathbb{V}}$ in Eq.~\eqref{mastereqeven}, and  $\vec{v}_{+,(0)}$ are the corresponding eigenvectors. To the leading orders, we have
\begin{align}
    \label{poteveng}
    V^{+}&=V^{+}_{\mathrm{GR}}+\alpha^{2} c_1^{2} \delta V^{+}_{1},\\
    \label{potevens}
    V^{(0)}&=V^{(0)}_{\mathrm{GR}}+\alpha c_2 \delta V_{2}^{(0)}+\alpha^{2} c_1^{2} \delta V_1^{(0)},
\end{align} 
where $\delta V^{+,(0)}_{1}$ and $\delta V_2^{(0)}$ denote the sGB corrections, the explicit expressions of which can be found in Appendix \ref{app:Bg&mastereqsdetails}. The quantities $r_{*\mathrm{T}}^{\pm,(0)}$ are the turning points defined by $\omega^{2}-V^{\pm,(0)}=0$. When $r_* \gtrless r_{*\mathrm{T}}^{\pm,(0)} $, we have $\omega^{2}-V^{\pm,(0)}\gtrless 0$. Using the WKB connected formula, we can infer the asymptotic scattering solutions at spatial infinity ($r_* \rightarrow \infty$), which consist of incident and reflected waves,  as
 \begin{align}
\Psi_{\omega \ell}^{-} \propto   e^{-i \omega r_*} +  e^{2 i \delta_{\omega\ell}^{-}} e^{i \omega r_*},
 \end{align}
 and
 \begin{align}
    \begin{pmatrix} \tilde{\Psi}_{\omega \ell}^{+} \\ \addlinespace \tilde{\Phi}_{\omega \ell}  \\\end{pmatrix} \propto \sum_{(k)=+,(0)}\vec{\tilde{v}}_{(k)}\left(e^{-i \omega r_{*}} + e^{2 i \delta_{\omega \ell}^{(k)}} e ^{i \omega r_{*}}\right).
 \end{align}
Here,  $2 \delta_{\omega\ell}^{\pm,(0)}$ are the scattering phase shifts for different modes, which are given by 
 \begin{align}
    \label{phaseshift}
     \delta_{\omega\ell}^{\pm,(0)} \simeq &\int_{r_{*\mathrm{T}}^{\pm,(0)}}^{\infty} \mathrm{d} r_*\left(\sqrt{\omega^2-V^{\pm,(0)}}-\omega\right) -\omega r_{*\mathrm{T}}^{\pm,(0)}-\frac{\pi}{4} ,
  \end{align}
and $\vec{\tilde{v}}_{+,(0)}$ are two normalized and orthogonal vectors defined by the asymptotic values of  $\vec{v}_{+,(0)}$. For the even modes, considering the transformation $(\Psi_{\omega \ell}^{+},\Phi_{\omega \ell} )^T = U (\tilde{\Psi}^{+}_{\omega \ell},\tilde{\Phi}_{\omega \ell} )^T$,  where 
$U=(\vec{\tilde{v}}_{+},\vec{\tilde{v}}_{(0)})^{-1}$ is a $2 \times 2 $ matrix, we obtain the asymptotic solutions at spatial infinity as
\begin{align}
    \begin{pmatrix} \Psi_{\omega \ell}^{+} \\ \addlinespace \Phi_{\omega \ell}  \\\end{pmatrix} \propto \begin{pmatrix} e^{-i \omega r_{*}} + e^{2 i \delta_{\omega \ell}^{+}} e ^{i \omega r_{*}} \\ e^{-i \omega r_{*}} + e^{2 i \delta_{\omega \ell}^{(0)}} e ^{i \omega r_{*}} \end{pmatrix}.
\end{align} 
In the following, we shall refer to $\Psi_{\omega \ell}^{\pm}$ as spin-$2$ modes and $\Phi_{\omega \ell}$ as scalar modes.

\begin{figure*}[tp]
    \centering
		 \includegraphics[width=0.44\textwidth]{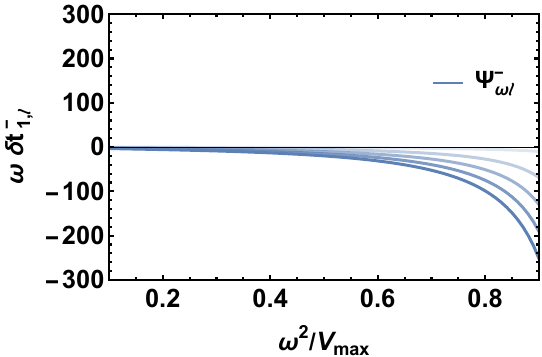}
         \qquad \qquad
         \includegraphics[width=0.45\textwidth]{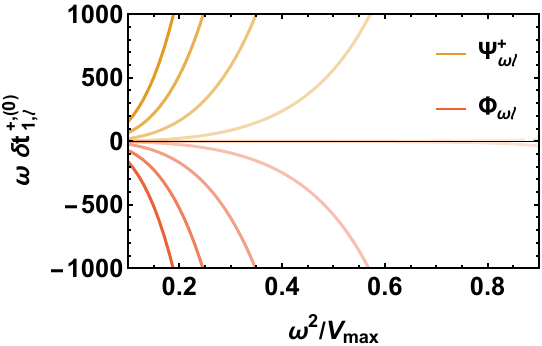}
          \qquad \qquad
		 \includegraphics[width=0.44\textwidth]{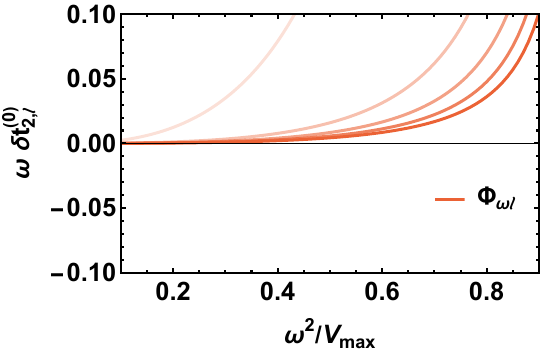}
    \caption{sGB corrections on the time delays for modes of different frequency $\omega$ and degree $\ell$. Here $V_{\rm max}$ is the maximum of the corresponding GR potentials, and curves, from light to dark, represent modes with  $\ell=2,22,42,62$ and $82$. The top-left and top-right panels show the corrections from the $c_1$-term on the odd and even modes, respectively, and the bottom panel shows those from the $c_2$-term. As in Eqs.~\eqref{mastereqodd} and~\eqref{mastereqeven}, the $c_2$-term only affects the scalar modes.}
    \label{fig:dtwrelation}
\end{figure*}

Given the phase shifts, we can further define the time delays for the spin-$2$ and scalar modes as
\begin{align}
    \label{asymptoticacausal}
    \Delta T_{\ell}^{\pm ,(0)}=2\frac{\partial \delta^{\pm ,(0)} _{\omega\ell}}{\partial \omega},
\end{align}
which, in GR, are known as the Eisenbud-Wigner time delays~\cite{Wigner:1955zz}. According to the prescription of infrared causality~\cite{deRham:1909.00881Speed,deRham:2007.01847Causality}, we are interested in extracting the time delay arising from the corrections from the sGB EFT couplings, which is the total time delay subtracted by the GR part,
\begin{equation}
   \Delta  T^{\pm ,(0)} _{\ell,\mathrm{sGB}}=\Delta T^{\pm ,(0)}_{\ell}-\Delta T^{\pm ,(0)}_{\ell,\mathrm{GR}}.
\end{equation}
Here, $\Delta T_{\ell,\mathrm{GR}}^{\pm (0)}$ are the GR time delays for the spin-$2$ and scalar modes, which are given by
\begin{align}
    \Delta T_{\ell,\mathrm{GR}}^{\pm, (0)} &= 2 \int_{r_{\mathrm{T,GR}}^{\pm, (0)}}^{\infty} \frac{\mathrm{d} r}{A_0}\left(\frac{2 \omega}{2 \sqrt{\omega^2-V_{\mathrm{GR}}^{\pm,(0)}}}-1\right) -2 r_{*\mathrm{T,GR}}^{\pm,(0)}.
\end{align}
Here, $r_{\mathrm{T,GR}}^{\pm, (0)}$ are the GR turning points defined by $\omega^{2}-V_{\mathrm{GR}}^{\pm,(0)}=0$ and $r_{*\mathrm{T,GR}}^{\pm,(0)}$ are the GR tortoise coordinate. To the leading orders, we have
\begin{align}
    \label{DTg}
    \Delta  T^{\pm } _{\ell,\mathrm{sGB}} &\simeq \alpha^2 c_1^2 \delta t_{1,\ell}^{\pm },\\
    \label{DTs}
    \Delta  T^{(0)}_{\ell,\mathrm{sGB}} &\simeq \alpha c_2 \delta t_{2,\ell}^{(0)}+\alpha^2 c_1^2 \delta t_{1,\ell}^{(0)},
\end{align}
where $\delta t_{1,\ell}^{\pm ,(0)}$ denote the leading-order corrections contributed by the $c_1$-term, and $\delta t_{2,\ell}^{(0)}$ denote the leading-order correction contributed by the $c_2$-term. Note that the corrections to the spin-$2$ modes are only affected by the $c_1$-term, and when $c_1=0$, the BH background will not be modified perturbatively and only the scalar modes will suffer corrections. According to Eqs.~\eqref{phaseshift} and~\eqref{asymptoticacausal}, we can analytically express $\delta t_{1,\ell}^{\pm ,(0)}$ and $\delta t_{2,\ell}^{(0)}$ in terms of integrals~\cite{deRham:2112.05054Causality,deRham:2007.01847Causality}, 
 \begin{align}
   \label{timedelayLO1}
   \delta t_{1,\ell}^{\pm, (0)} &=-2 \int_{r_{*\mathrm{T,GR}}^{\pm,(0)}}^{\infty} \mathrm{d} r \mathcal{A}^{\pm, (0)} \left(\frac{\delta \mathcal{A}^{\pm,(0)}_1}{\mathcal{A}^{\pm,(0)\prime}}\right)^{\prime},\\
   \label{timedelayLO2}
   \delta t_{2,\ell}^{(0)}&=-2 \int_{r_{*\mathrm{T,GR}}^{(0)}}^{\infty} \mathrm{d} r \mathcal{A}^{(0)} \left(\frac{\delta \mathcal{A}^{(0)}_2}{\mathcal{A}^{(0)\prime}}\right)^{\prime},
 \end{align}
where 
\begin{align}
    \mathcal{A}^{\pm, (0)}  & \equiv  \frac{\omega}{A_0 \sqrt{\omega^2-V_{\mathrm{GR}}^{\pm,(0)}}}, \\
   \delta \mathcal{A}_1^{\pm,(0)}  & \equiv  \mathcal{A}^{\pm,(0)}\bigg[\frac{\delta V_{1}^{\pm,(0)}}{2\left(\omega^2-V_{\mathrm{GR}}^{\pm,(0)} \right)}    -\frac{1}{2 \omega} \frac{\partial \delta V_1 ^{\pm,(0)}}{\partial \omega}  -\frac{\delta A}{A_0}\bigg], \\
    \delta \mathcal{A}_2^{(0)} & \equiv  \mathcal{A}^{(0)}\frac{\delta V_{2}^{(0)}}{2\left(\omega^2-V_{\mathrm{GR}}^{(0)}\right)},
\end{align}
with $\delta A$ defined as
\begin{align}
  c_1^2 \alpha^2 \delta A & \equiv  \sqrt{(A_0+\alpha^2 c_1^2 A_1) (B_0+\alpha^2 c_1^2 B_1)}-A_0 +\mathcal{O}(\alpha^4).
\end{align}
By numerical integration, we can get the values of $\delta t_{1,\ell}^{\pm ,(0)}$ and $\delta t_{2,\ell}^{(0)}$. In Fig.~\ref{fig:dtwrelation}, we show the numerical results of the sGB corrections on the time delays for both spin-$2$  and scalar modes at different frequency $\omega$ and degree $\ell$. 

It is instructive to analytically estimate the $\ell$ scalings for various quantities involved at fixed $\gamma$ and large $\ell$. By the explicit expressions for the potentials in Appendix \ref{app:Bg&mastereqsdetails}, we find that, in the large-$\ell$ limit, $V_{\mathrm{max}}$, $V_{\mathrm{GR}}^{\pm, (0)}$, and $\delta V_{1}^{-}$ scale as $\ell^2$, $\delta V_{1}^{+,(0)}$ scales as $\ell^4$, and $\delta V_{2}^{(0)}$ scales as $\ell^0$. Consequently, we have $\omega \sim \ell$ and $r_{\mathrm{T,GR}}^{\pm, (0)}$ are independent of $\ell$. Then, from Eqs.~\eqref{timedelayLO1} and \eqref{timedelayLO2}, we deduce that $\delta t_{1,\ell}^{-} \sim \ell^0$, $\delta t_{1,\ell}^{+ ,(0)} \sim \ell^2$, and $\delta t_{2,\ell}^{(0)} \sim \ell^{-2}$.

When $\omega^{2}>V_{\mathrm{max}}$, incident waves will enter the BH horizon instead of being scattered back to spatial infinity, so we can not use the phase shift of the outgoing wave to define time delay. In this case, let us consider minimally coupled photons and gravitational/scalar waves that travel radially outwards from a radius $r_0$ near the BH horizon to spatial infinity. We can define a new type of time delay by comparing the travel time between them. This results in the following time advance for the GWs, as compared to the photons~\cite{deRham:2007.01847Causality},
\begin{align}
\label{tmadvlargeo}
  \Delta T ^{\pm}_{\mathrm{adv}} \sim \int_{r_0}^{\infty} \frac{\mathrm{d} r \Delta c_s^{ \pm }}{2\left(1-\frac{r_g}{r}\right)},
\end{align}
where $\Delta c_s^{ \pm }$ represent the leading-order sGB corrections on the radial sound speeds of the GWs, and we have taken the limit  $r_0 \rightarrow r_g$. See Appendix~\ref{app:Bg&mastereqsdetails} for the explicit expressions of $\Delta c_s^{ \pm }$. For the scalar waves, since there is no correction on their sound speeds at leading order (see Appendix~\ref{app:Bg&mastereqsdetails}), we don't have to consider their time advance.  Performing the integrations in Eq.~\eqref{tmadvlargeo} yields $\Delta T ^{\pm}_{\mathrm{adv}}<0$, indicating that there is no time advance in this case. Therefore, causality is automatically respected.

\begin{figure*}[tp]
   \centering
   \includegraphics[width=0.44\textwidth]{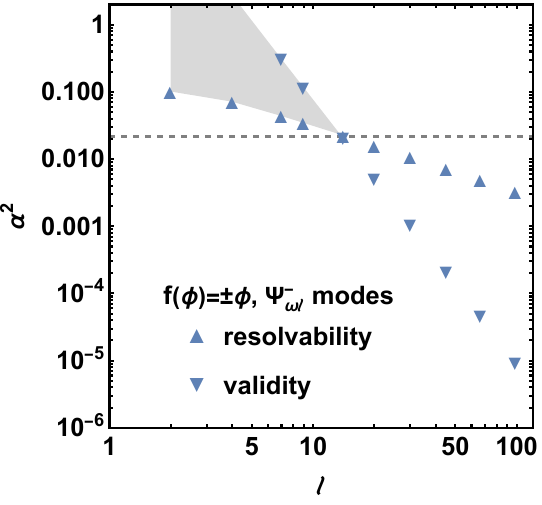} \qquad \qquad
    \includegraphics[width=0.45\textwidth]{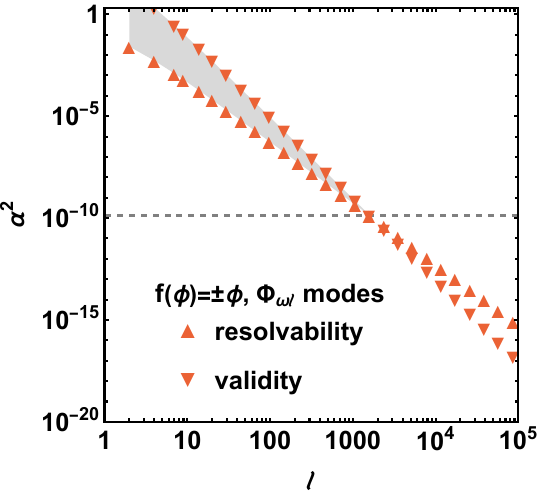}
   \includegraphics[width=0.44\textwidth]{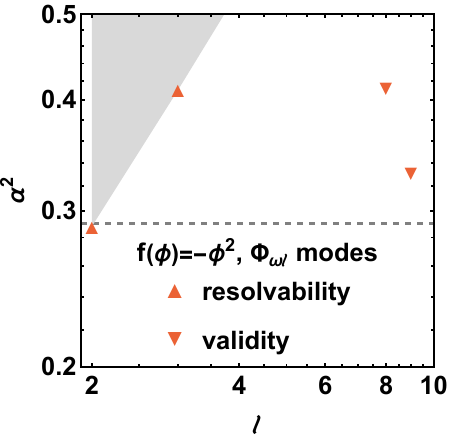}
    \caption{Causality violating regions (shadowed regions) in sGB gravity. The top-left and top-right panels show the parameter spaces for the odd spin-$2$ modes $\Psi^{-}_{\omega \ell}$ and the scalar modes $\Phi_{\omega \ell}$, respectively, for the case of $|c_1| = 1$ and $c_2=0$, and the bottom panel shows the parameter space for $\Phi_{\omega \ell}$ for the case of $c_1=0$ and $c_2 = -1$. A solid, up-pointing triangle denotes the resolvability condition for each $\ell$, while a solid, down-pointing triangle denotes the validity condition. The dashed lines denote the upper bounds on $\alpha^2$ or $\alpha$ imposed by infrared causality. }
    \label{fig:esgbalphabound}
\end{figure*}

\section{Causality constraints}
\label{sec:causal}

\subsection{sGB gravity}

As shown in Fig.~\ref{fig:dtwrelation}, with some choices of $c_1$ and $c_2$, the sGB corrections on the time delay could be negative, i.e., $\Delta T^{(0)}_{\ell,\mathrm{sGB}}<0$. As stated previously in the introduction, a negative $\Delta T^{(0)}_{\ell,\mathrm{sGB}}$ indicates that the low energy modes, in our case the scattering waves, propagate outside the light cone set by the high energy modes. In particular, if the negative corrections on the time delay is resolvable\footnote{The resolvability condition arises from the uncertainty principle. For a wavepacket of energy $\omega$, the uncertainty in the time delay (or time advance) is around $\omega^{-1} $. If the time advance does not exceed this uncertainty, causality can still be respected.} ,  i.e., $-\Delta T^{(0)}_{\ell,\mathrm{sGB}} > \omega^{-1}$, it would imply violation of infrared causality~\cite{deRham:1909.00881Speed,deRham:2007.01847Causality}.

To be concrete, let us first consider sGB gravity with $c_1 = 0$ and $c_2 \neq 0$, in which case the sGB coupling only affects the scalar modes up to $\mathcal{O}(\alpha)$, and $\Delta T^{(0)}_{\ell,\mathrm{sGB}}$ is negative whenever $c_2 < 0$. To carve out the boundary of the causality bounds, it is convenient to absorb $|c_2|$ in $\Lambda$ by setting $c_2=-1$. Then, the resolvability of $\Delta T^{(0)}_{\ell,\mathrm{sGB}}$ requires $\alpha \delta t_{2, \ell}^{(0)} < \omega^{-1}$. On the other hand, to ensure the EFT validity, both the BH background and the scattering waves must be under control within the sGB EFT. In other words, all possible Lorentz invariant (background and perturbative) quantities constructed in the scattering setup should be controlled by the cutoff scale $\Lambda$ \cite{deRham:2007.01847Causality}. More specifically, for a scalar-tensor EFT, the validity requirement enforces the following conditions:
\begin{align}
\label{nnRRconditions}
(\bar{\nabla}_{\mu}\bar{\nabla}^{\mu})^m (k_1^{\mu}k_{2\mu})^n  (\bar{R}_{\mu \nu \rho \sigma}\bar{R}^{\mu \nu \rho \sigma})^p \bar{\phi}^q  \ll \Lambda^{2m+2n+4p} ,
\end{align}
where $\bar{\nabla}_{\mu}$ is the covariant derivative associated with the background, $\bar{R}_{\mu \nu \rho \sigma}$ and $\bar{\phi}$ denote the background Riemann tensor and scalar field respectively, and $k_{1\mu}=(-\omega,\omega /\sqrt{A B} \hat{k}_{1})$ and $k_{2\mu}=(-\omega,\omega /\sqrt{A B} \hat{k}_{2})$ are the momenta of the on-shell GW and scalar wave respectively ($\hat{k}_{1}$ and $\hat{k}_{2}$ being two unit $3$-vectors). To transform the conditions \eqref{nnRRconditions} into forms that are readily implementable, first note that by special choices of $m,n,p$ and $q$, we have  natural requirements $\bar{\nabla}_{\mu}\bar{\nabla}^{\mu} \ll \Lambda^2$,  $k_1^{\mu}k_{2\mu} \ll \Lambda^2$,  $\bar{R}_{\mu \nu \rho \sigma}\bar{R}^{\mu \nu \rho \sigma}\ll \Lambda^4$ and $\bar{\phi} \ll 1$. With these conditions, it is easy to see that the strongest conditions from \eqref{nnRRconditions} can be obtained by taking $m\to \infty$, $n \rightarrow \infty$, $p \rightarrow \infty$ and $q \rightarrow \infty$ separately while keeping the rest fixed, which again gives the above four conditions. Also note that, for our background solution, we have the following scalings $k_{1,2\mu} \sim \omega$, $\bar{R}_{\mu \nu \rho \sigma}\sim GM/r^3$, $\bar{\phi}\sim \alpha G M /r $, and $\bar{\nabla}_{\mu} \sim 1/r$. Thus, the strongest validity condition from the background is $GM \gg \Lambda^{-1}$, while the strongest one from waves is from the case with $\hat{k}_{1} =- \hat{k}_{2}$, which gives $\omega \ll \Lambda$.   
Taking all of these into consideration, it turns out that, for $\alpha \ll 1$, the strongest EFT validity condition is
\begin{align}
    \alpha \ll  \frac{1}{\omega^{2} G^{2} M^{2} } . 
\end{align}
In Fig.~\ref{fig:esgbalphabound}, we show the constraints on $\alpha$ imposed by infrared causality, where an up-pointing triangle denotes the resolvability condition for each $\ell$, a down-pointing triangle denotes the validity condition, and the shadowed regions violate infrared causality. We consider waves with $\gamma = 0.9$ such that we can get relatively tight constraints (cf.~Fig.~\ref{fig:dtwrelation}). From Fig.~\ref{fig:esgbalphabound}, we can conclude that infrared causality requires $\alpha \lesssim 0.3$ for sGB gravity with $c_1=0$ and $c_2 = -1$. 

Let us now consider sGB gravity with $c_1 \neq 0$, in which case, the sGB coupling always leads to negative corrections on the time delays for the $\Psi^{-}_{\omega \ell}$ modes and positive corrections for the $\Psi^{+}_{\omega \ell}$ modes; cf.~Fig.~\ref{fig:dtwrelation} and Eq.~\eqref{DTg}. Therefore, for the $\Psi^{+}_{\omega \ell}$ modes, causality is always respected and no constraints arise. Moreover, for the scalar modes, both the $c_1$- and $c_2$-term contribute, and it seems that a negative $c_2$ could improve the causality bound derived from the scalar modes. However, assuming $c_1$ and $c_2$ to be ${\cal O}(1)$, the $c_2$-term does not affect the causality constraint very much due to a relatively small $\delta t_{2,\ell}^{(0)}$. So infrared acausality imposes
\begin{align}
   \label{IRboundalpha}
    \frac{1}{- \omega \delta t_{1,\ell}^{-,(0)}} \lesssim c_1^2  \alpha^2 \ll \bigg(\frac{1}{\omega ^2 G^2 M^2}\bigg)^2  ,
\end{align}
where the lower bound is the resolvability condition, and the upper bound is the EFT validity condition. In the large-$\ell$ limit, the upper bound scales as $\ell^{-4}$, while the lower bound scales as $\ell^{-1}$ for the odd modes and $\ell^{-3}$ for scalar modes. Numerically, the two-sided bounds for various $\ell$ modes can be found in Fig.~\ref{fig:esgbalphabound}, which gives the bound
\begin{align}
    \alpha \lesssim 1.1 \times 10^{-5},
\end{align}
namely 
\begin{align}
    \Lambda \gtrsim 1.3 \times 10^{-8} \left(\frac{3 M_{\odot}}{M}\right) \mathrm{eV},
\end{align}
with $M_{\odot}$ being the mass of the sun, for sGB gravity with $c_1 =\pm 1$.

\subsection{Including the cubic curvature operator}

\begin{figure*}[tp]
    \centering
		 \includegraphics[width=0.44\textwidth]{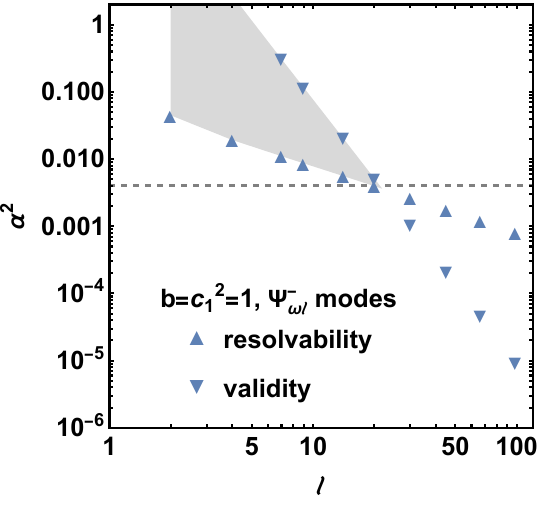}\qquad \qquad
		 \includegraphics[width=0.44\textwidth]{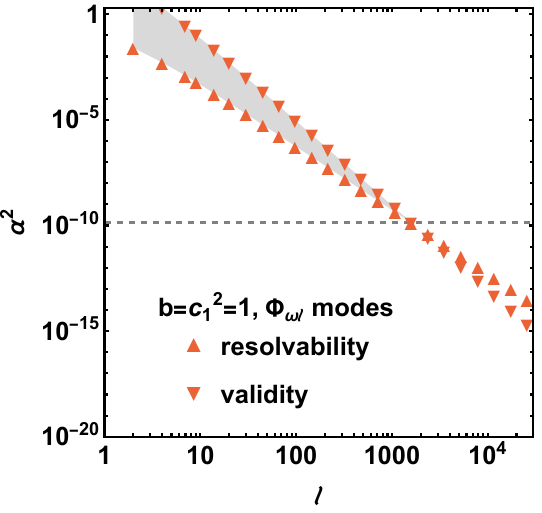}
    \caption{Causality violating regions (shadowed regions) in sGB gravity with the cubic curvature operator included. The dashed lines denote the upper bounds on $\alpha^2$.}
    \label{fig:dim6sGB}
\end{figure*}

For more generic EFTs, there are also higher-dimensional operators from graviton self-interactions. In this subsection, we discuss the effects of further including these operators on the causality bounds. As discussed in Ref.~\cite{deRham:2112.05054Causality}, the leading-order such operator is the cubic curvature term $R_{\mu \nu}{ }^{\alpha \beta} R_{\alpha \beta}{ }^{\gamma \sigma} R_{\gamma \sigma}{ }^{\mu \nu}$, which contributes to the scattering time delay at $\mathcal{O}(\alpha^2)$, the same order as the linear sGB coupling. To be concrete, the action we consider in this subsection is 
\begin{align}
    S&= \frac{M_{\mathrm{Pl}}^2}{2} \int \mathrm{~d}^4 x \sqrt{-g}\bigg[R-\frac{1}{2} \partial_\mu \phi \partial^\mu \phi  +  \frac{c_1}{\Lambda^2} \phi \mathcal{G} +\frac{b}{\Lambda^4} R_{\mu \nu}{ }^{\alpha \beta} R_{\alpha \beta}{ }^{\gamma \sigma} R_{\gamma \sigma}{ }^{\mu \nu}\bigg],
\end{align}
where $b$ is also a dimensionless coupling constant. The steps to calculate the time delays are very similar to those for the case with only the sGB term, so they will not be repeated here. If $b$ and $c_1$ are both $\mathcal{O}(1)$, these two terms can lead to comparable corrections on the time delays for the odd modes, while for the even modes, the corrections from the $b$-term are relatively small compared to those from the $c_1$ term. In Fig.~\ref{fig:dim6sGB}, we plot the results when both the sGB term and the cubic curvature term are included, setting $b=c_1^2=1$ for simplicity. We see that the causality bound remains around $\alpha \lesssim 1.1 \times 10^{-5}$. Thus, including the cubic curvature operator has a negligible effect on the bound on the bound obtained for sGB gravity above.

\subsection{Including multi-sGB couplings}

\begin{figure}[tp]
    \centering
    \includegraphics[width=0.44\textwidth]{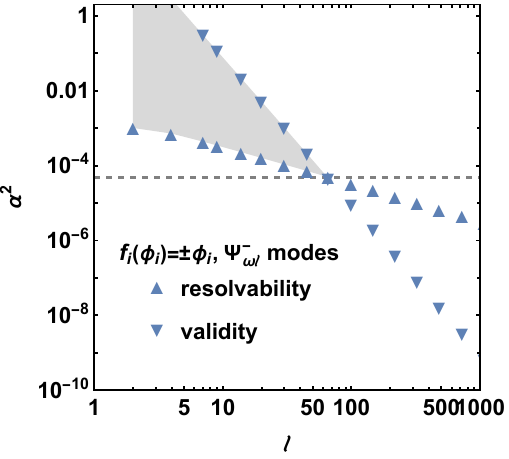}\qquad\qquad
     \includegraphics[width=0.43\textwidth]{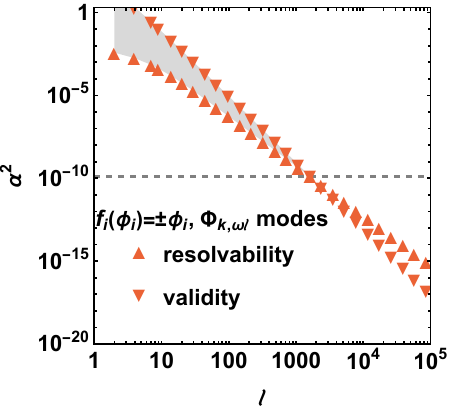}\,
     \caption{Causality violating regions (shadowed regions) in MsGB gravity with $N=100$ scalar fields ($|c_{k,1}| = 1$ and $c_{k,2}=0$). The dashed lines denote the upper bounds on $\alpha^2$ imposed by infrared causality.}
     \label{fig:mesgbalphabound}
\end{figure}

 In some fundamental theories, scalars are ubiquitous, and there could be many scalars that become relevant in the strong gravity environment. So let us also entertain the possibility of multi-sGB (MsGB) couplings. Analogous to Eq.~\eqref{ESGBAction}, the action for MsGB gravity with $N$ scalar degrees of freedom can be written as 
\begin{align}
    \label{MESGBAction}
    S&= \frac{M_{\mathrm{Pl}}^2}{2}  \int \mathrm{d}^4 x \sqrt{-g} \bigg\{R + \sum_{k=1}^{N} \left[-\frac{1}{2}\partial _\mu \phi_{k}\partial ^\mu \phi_{k}+\frac{1}{\Lambda^2} f_{k}(\phi_{k}) \mathcal{G}\right]\bigg\}  ,
\end{align}
where $\phi_{k}$ is the $k$-th scalar field, and $f_{k}$ is the $k$-th coupling function, expanded as 
\begin{align}
    f_{k}(\phi)=c_{k,1} \phi_{k} + c_{k,2} \phi_{k}^2 + \cdots ,
\end{align}
where $c_{k,1}$ and $c_{k,2}$ are the coupling constants and the dots represent higher-order terms. The steps to calculate the time delays are very similar to those for the case with only the sGB term, so they will not be repeated here. Our result is as follows. For the spin-$2$ modes $\Psi_{\omega \ell}^{\pm}$, the MsGB corrections on their time delays at leading order are given by 
\begin{align}
    \Delta  T^{\pm } _{\ell,\mathrm{sGB}} \simeq & \alpha^2 \sum_{k=1}^{N} c_{k,1}^2 \delta t_{1,\ell}^{\pm },
\end{align}
while, for the $k$-th scalar modes $\Phi_{k,\omega \ell}$, the leading-order corrections are given by
\begin{align}
    \Delta  T^{(0)}_{k,\ell,\mathrm{sGB}} \simeq & \alpha c_{k,2} \delta t_{2,\ell}^{(0)}+\alpha^2 c_{k,1}^2 \delta t_{1,\ell}^{(0)} ,
\end{align}
where $k=1,2, \ldots N$. Compared to Eqs.~\eqref{DTg} and~\eqref{DTs}, adding more scalar degrees of freedom affects only the time delay for the spin-$2$ modes through the coupling constants in front of $\delta t_{1,\ell}^{\pm}$. Therefore, only the causality bounds on the spin-$2$ could become tighter, while the causality bounds on the scalar modes remain the same. In Fig.~\ref{fig:mesgbalphabound}, we consider the case $N=100$ and set $c_{k,1}=1$ for simplicity. We find that in this case, even the causality bounds on the spin-$2$ modes become tighter compared with Fig.~\ref{fig:esgbalphabound}, it is still weaker than the causality bounds determined by the scalar modes. Therefore, the causality bound remains $\alpha \lesssim 1.1 \times 10^{-5}$, which is the same as in sGB gravity.

\section{Detectability of sGB gravity}
\label{sec:detect}

\begin{figure*}[tp]
    \centering
     \includegraphics[width=0.44\linewidth]{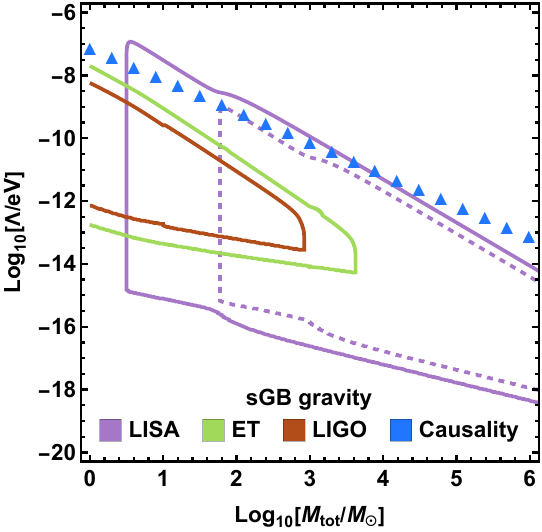}
     \qquad\qquad
    \includegraphics[width=0.44\linewidth]{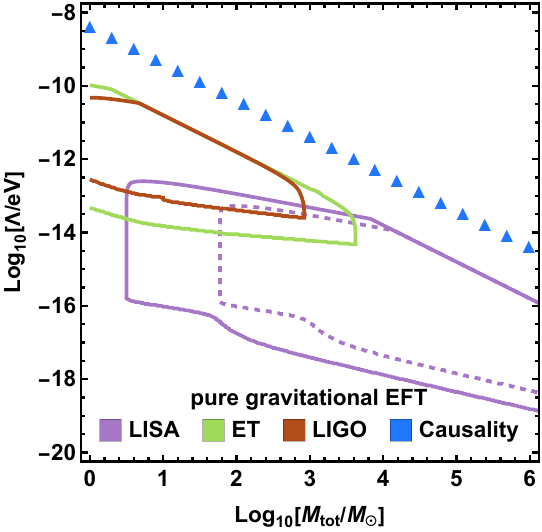}
    \caption{Detectability of sGB gravity (left panel) and pure gravitational EFT (right panel) using inspiral GWs from equal-mass binary BHs. $\Lambda$ is the EFT cutoff, and $M_{\rm tot}$ is the total mass of the binary. A blue, up-pointing triangle represents the lower bound on $\Lambda$ imposed by infrared causality for a given $M_{\mathrm{tot}}$. We consider binaries inspiraling at 300 Mpc for LIGO, at 300 Mpc for Einstein Telescope (ET), and at 300 Mpc (solid line) or 3 Gpc (dashed line) for LISA. Detection becomes possible when the causality lower bound enters the GW sensitivity curves.}
\label{fig:ESGBGWObs}
\end{figure*}

With rapid advances in GW astronomy, we are interested in whether we can test consistent gravitational EFTs with GW observations, taking into account the causality constraints derived above. Before discussing the observability of sGB gravity, we first briefly review the interplay between causality bounds and GW experiments for the pure gravitational EFT (without the scalar field)~\cite{deRham:2112.05054Causality}. As pointed out in Ref.~\cite{deRham:2112.05054Causality}, causality constraints indicate that the cubic curvature operator $R_{\mu \nu}{ }^{\alpha \beta} R_{\alpha \beta}{ }^{\gamma \sigma} R_{\gamma \sigma}{ }^{\mu \nu}$ can not be tested with GW signals in the near future. For the ringdown tests, the leading-order beyond-Einstein correction, e.g., on the BH quasi-normal modes are proportional to $(GM\Lambda)^{-4}$, which has to be small for the EFT to be valid on the BH horizon scale. Therefore, this cubic curvature operator can hardly be tested with the BH ringdown waveforms given the precision of current GW observations. Nevertheless, the EFT with a lower cutoff might still be tested with early inspirals. For an EFT cutoff $\Lambda$, one can consider a period of inspiral with an orbital separation larger than $R_{\rm min}$. In this case, $R_{\mathrm{min}}$ is usually larger than the radius of the innermost stable circular orbit. Although the EFT corrections from the cubic curvature operator on the inspiral waveform are still proportional to $(GM_{\mathrm{tot}}\Lambda)^{-4}$, where $M_{\mathrm{tot}}$ is the total mass of the binary, it only requires $(\Lambda R_{\rm min})^{-1} \ll 1$ for the EFT to be valid for describing such a period of inspiral.\footnote{Strictly speaking, for pure gravitational EFTs, the EFT validity requires $f < \Lambda^2 R_{\rm min}$~\cite{Chen:2112.05031causality}, which is slightly stronger than $(\Lambda R_{\rm min})^{-1} \ll 1$.} $(GM_{\mathrm{tot}}\Lambda)^{-4}$ can still be considerable given that $R_{\rm min}/GM_{\mathrm{tot}}$ is large in early inspirals. However, causality imposes another condition besides the EFT validity. In particular, by considering a fiducial black hole of mass $R_{\rm min}/G$, causality requires $(\Lambda R_{\rm min})^{-4} < 1.3 \times 10^{-5}$ for the pure gravitational EFT.  As shown in the right panel of Fig.~\ref{fig:ESGBGWObs}, such causality constraints are so strong that they eliminate all of the parameter space for testing the pure gravitational EFT with early inspirals.

Now, we consider testing sGB gravity with binary BH inspirals. The monopole charge of the hairy BH can lead to the emission of scalar dipole radiation. As a result, for a binary BH with mass $M_1$ and $M_2$ for the two BHs, in the Post-Newtonian (PN) expansions, the beyond-Einstein correction on the GW phase of the binary BH inspiral, $\delta \psi_{\mathrm{sGB} }$, starts at $-1$PN relative order and is given by~\cite{Yagi:1110.5950Inspirals}
\begin{align}
    \label{phasesgb}
    \delta \psi_{\mathrm{sGB} }(f) = \beta_{\mathrm{sGB}} 
    \left(G M_{\text{tot}}\Lambda \right)^{-4}  \left(G \mathcal{M} \pi f\right)^{-\frac{7}{3}} ,
\end{align}
where $f$ is the GW frequency, $\mathcal{M}=(M_1 M_2)^{3/5} M_{\text{tot}}^{-1/5}$ is the chirp mass, and $\beta_{\mathrm{sGB}}$ is a dimensionless coefficient given by~\cite{Berti:2018vdi} 
\begin{align}
    \beta_{\mathrm{sGB}}=-\frac{5}{7168}  \frac{\left(M_1^2 \tilde{s}_2^{\mathrm{sGB}}-M_2^2 \tilde{s}_1^{\mathrm{sGB}}\right)^2}{M_{\text{tot}}^4 \eta^{18 / 5}},
\end{align}
where $\eta=M_1 M_2/M_{\text{tot}}^2$ is the symmetric mass ratio, and $\tilde{s}_{i}^{\mathrm{sGB}}= 2(\sqrt{1-\chi_i^2}-1+\chi_i^2) / \chi_i^2$ are the dimensionless factors of the scalar monopole charges, with $\chi_i=\vec{S}_i \cdot \hat{L} / M_i^2$ being the projection of the BH spin angular momentum $\vec{S}_i$ in the direction of the orbital angular momentum $\hat{L} $. To estimate the observability of the sGB corrections, we calculate the accumulated phase shift $\Delta \psi_{\mathrm{sGB}}$ caused by the sGB couplings during the inspiral, 
\begin{align}
    \Delta \psi_{\mathrm{sGB}}=\left|\delta \psi_{\mathrm{sGB}}\left(f_{ \mathrm{max}}\right)-\delta \psi_{\mathrm{sGB}}\left(f_{  \mathrm{min}}\right)\right|,
\end{align}
where $f_{\mathrm{max}}$ and $f_{\mathrm{min}}$ are the maximal and minimal frequencies of the inspiral signal. The inspiral frequency band $[f_{\mathrm{min}},f_{\mathrm{max}}]$ is determined by two aspects: the band depends on the capability of the  GW detector, and the frequency should exceed neither the cutoff frequency\footnote{Here, the EFT validity requires $(\Lambda R_{\mathrm{min}})^{-1}<1$. According to Kepler's third law, we have $R_{\mathrm{min}}^3=(\pi f_{\mathrm{cut}})^{-2} G M_{\text{tot}}$, where $f_{\mathrm{cut}}$ is the cutoff frequency. Therefore, we get $f_{\mathrm{cut}}= \pi^{-1}(G M_{\mathrm{tot}} \Lambda^{3})^{-2}$.} nor the innermost stable circular orbit frequency. Since the charge factors $\tilde{s}_{i}^{\mathrm{sGB}}$ must be within the range $0\leq \tilde{s}_{i}^{\mathrm{sGB}}\leq 1$, in the following discussions, we focus on the case when $\tilde{s}_{1}^{\mathrm{sGB}}=1$ and $\tilde{s}_{2}^{\mathrm{sGB}}=0$ for relatively large phase shifts. The results are shown in the left panel of Fig.~\ref{fig:ESGBGWObs}, where the contours show  $\Delta \psi_{\mathrm{sGB}} \gtrsim 1$, indicating that sGB gravity can be tested within the parameter regime bounded by the contours \cite{Sennett:1912.09917GREFTGW,deRham:2112.05054Causality}. When $\Delta \psi_{\mathrm{sGB}} < 1$, we cannot distinguish between sGB gravity and GR when analyzing inspiral waveforms in GW experiments. On the other hand, we also plot the causality constraints, shown by the blue, up-pointing triangles. As calculating GW waveforms in sGB gravity requires knowledge of the BH solution, as in Ref.~\cite{Yagi:1110.5950Inspirals}, the theory should be at least causal down to the scale of BH horizon, i.e., $(\Lambda G M_{\rm tot}/2 )^{-1} < 0.003$.\footnote{This is different from the pure gravitational EFTs discussed in Ref.~\cite{deRham:2112.05054Causality}, in which case the EFT only need to be valid and causal down to the scale of binary separation.} 

Due to the dipole radiation resulted from the scalar monopole hair, the sGB coupling typically leads to larger observational effects compared to the cubic curvature operator. As a result, unlike the pure gravitational EFT, sGB gravity can still be tested with GW inspirals, with causality constraints leaving a detectability window for this possibility. For example, if LISA is used to capture inspiral signals from binary BHs at a distance of 300 Mpc, it is possible to detect sGB gravity in the range $M_{\mathrm{tot}} \in [ 10, 10^3 ] \,M_{_{\odot}}$ or $\Lambda \in [ 10^{-10}, 10^{-7} ] \,\mathrm{eV}$.

\begin{figure*}[tp]
    \centering
     \includegraphics[width=0.44\linewidth]{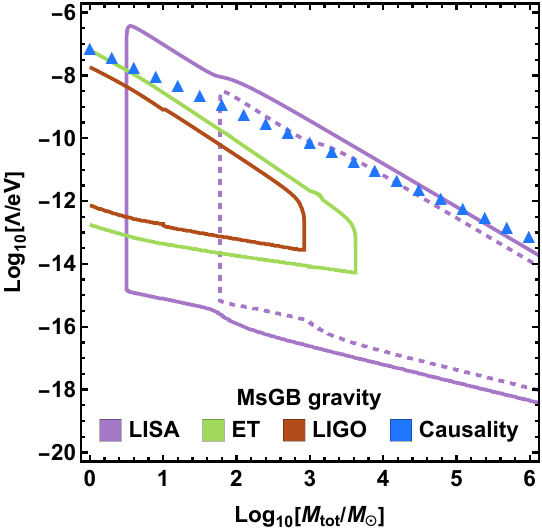}
    \caption{Detectability of MsGB gravity with $N=100$ scalar fields using inspiral GWs from equal-mass binary BHs. Here, we also consider binaries inspiraling at 300 Mpc for LIGO and Einstein Telescope (ET), and at 300 Mpc (solid line) or 3 Gpc (dashed line) for LISA, which are the same as in Fig.~\ref{fig:ESGBGWObs}.}
\label{fig:MESGBGWObs}
\end{figure*}

Finally, adding more scalar degrees of freedom can enlarge the detectability window, as the scalar dipole radiations can be enhanced in this case, leading to larger observational effects. For MsGB gravity (see Eq.~\eqref{MESGBAction}) with $N=100$, the leading-order MsGB correction on the GW phase of the binary BH inspiral is $\delta \psi_{\mathrm{MsGB} }(f)= N \delta \psi_{\mathrm{sGB} }(f) $. The discussions on the detectability of MsGB gravity are the same as for sGB gravity. Therefore, we present only the final results here. For example, as shown in Fig.~\ref{fig:MESGBGWObs}, for binary BHs at a distance of 300 Mpc, there is a detectable window with $M_{\mathrm{tot}} \in [ 10, 10^5 ]~ M_{_{\odot}} $ or $\Lambda \in [ 10^{-12}, 10^{-7} ] ~\mathrm{eV}$ with LISA. Similarly, for binary BHs at 3 Gpc, there is a detectable window with $M_{\mathrm{tot}} \in [ 10^{2}, 10^4 ] ~M_{_{\odot}} $ or $\Lambda \in [ 10^{-11}, 10^{-9} ] ~\mathrm{eV}$. Thus, the detectable window is enlarged.

\begin{acknowledgments}
We thank Claudia de Rham and Andrew J.~Tolley for helpful discussions. SYZ acknowledges support from the National Key R\&D Program of China under grant No.~2022YFC2204603 and from the National Natural Science Foundation of China under grants  No.~12475074, No.~12075233 and No.~12247103. JZ is supported by the scientific research starting grants from University of Chinese Academy of Sciences (grant No.~E4EQ6604X2), the Fundamental Research Funds for the Central Universities (grant No.~E2EG6602X2 and grant No.~E2ET0209X2) and the National Natural Science Foundation of China (NSFC) under grant No.~12147103 and grant No.~12347103.
\end{acknowledgments}

\appendix

\section{BHs and BH perturbations in sGB gravity}
\label{app:Bg&mastereqsdetails}

Using the static and spherically symmetric ansatz, and solving Eqs.~\eqref{ESGBEOM1} and~\eqref{ESGBEOM2} order by order in $\alpha$, we get 
\begin{align}
    A_1(x)&=\frac{1}{240 x^7}\left(10 x^4+130 x^3+33 x^2 +24 x-50\right), \\ 
    B_1(x)&=\frac{1}{240 x^7}\left(60 x^5+30 x^4+260 x^3+15 x^2+12 x-230 \right) ,\\
    \phi_1(x)&=\frac{1}{6 x^3} \left(6 x^2+3 x+2\right),\\
    \phi_2(x)&=\frac{1}{3600 x^6}\left(4380 x^5+2190 x^4+1460 x^3  +1095 x^2+336 x+100 \right),
\end{align}
where $x=r/r_g$ with $r_g=2GM$.

Next, we derive the master equations for metric and scalar perturbations on the BH background. To that end, we consider the metric perturbations $h_{\mu \nu}$
\begin{align}\label{meper}
    g_{\mu \nu}=\bar{g}_{\mu \nu}+h_{\mu \nu},
\end{align}
and the scalar perturbations $\delta \phi$ 
\begin{align}\label{sper}
    \phi=\bar{\phi}+\delta \phi,
\end{align}
where $\bar{g}_{\mu \nu}$ and $\bar{\phi}$ are the background determined by Eqs.~\eqref{background1}-\eqref{background3}. The perturbation equations are obtained by substituting Eqs.~\eqref{meper} and \eqref{sper} into Eqs.~\eqref{ESGBEOM1} and \eqref{ESGBEOM2}, and keeping the leading-orders of $h_{\mu \nu}$ and $\delta \phi$. The metric perturbations $h_{\mu \nu}$ can be split into  a sum of  odd-parity modes $h_{\mu \nu}^{-}$  and even-parity modes $h_{\mu \nu}^{+}$, while the scalar perturbations $\delta \phi$ are even-parity modes. We decompose the metric perturbations into tensor spherical harmonics in the frequency domain. Given the spherically symmetric background, there is no dependence on the order $m$, so we choose $m=0$ for simplicity. Also because of the spherical symmetry, modes of different degree $\ell$ decouple. In sGB gravity, since there is no parity-violation term, odd-parity modes and even-parity modes also decouple. In the Regge-Wheeler gauge, the metric perturbations of degree $\ell$ can be then written in matrix form as
\begin{align}
\label{mdeo}
h_{\mu \nu}^-=e^{-i \omega t}\left(\begin{array}{cccc}
0 & 0 & 0 & h_0 \\
0 & 0 & 0 & h_1 \\
0 & 0 & 0 & 0 \\
h_0 & h_1 & 0 & 0
\end{array}\right) \sin \theta Y_{\ell}^{\prime}(\theta),
\end{align}
and
\begin{align}
\label{mdee}
h_{\mu \nu}^{+}=e^{-i \omega t}\left(\begin{array}{cccc}
A H_0 & H_1 & 0 & 0 \\
H_1 & H_2 / B & 0 & 0 \\
0 & 0 & r^2 \mathcal{K} & 0 \\
0 & 0 & 0 & r^2 \sin ^2 \theta \mathcal{K}
\end{array}\right) Y_{\ell}(\theta) ,
\end{align}
while the scalar perturbations of degree $\ell$ can be written as
\begin{align}\label{sde}
    \delta \phi= e^{- i \omega t} \frac{\hat{\phi}}{r} Y_{\ell}(\theta),
\end{align}
where $Y_{\ell}(\theta)=Y_{\ell 0}(\theta, \phi)$ are the spherical harmonics with $m=0$ and $h_0, h_1, H_0, H_1, H_2,\mathcal{K}$ and $\hat{\phi}$ are functions of $r$. By substituting Eqs.~\eqref{mdeo},~\eqref{mdee} and~\eqref{sde} into the perturbation equations and considering the following definitions of master variables 
\begin{align}
    \Psi^{\mathrm{-}}_{\omega \ell} & =\bigg(1+\alpha^2 c_1^2 f^-_{h_1}\bigg)\frac{i \sqrt{A B} }{r \omega} h_1,
\end{align} 
and
\begin{equation}
  \begin{aligned}
    \begin{pmatrix} \tilde{\Psi}_{\omega \ell}^{+} \\ \addlinespace \addlinespace \tilde{\Phi}_{\omega \ell}   \\  \end{pmatrix} = 
    \begin{pmatrix} 1+\alpha^{2} c_1^{2} f_{\mathcal{K}} ^{+} & \alpha c_1 f^{+}_{\phi,1}+\alpha^{2} c_1 c_2 f^{+}_{\phi,2}  \\ \addlinespace \addlinespace \alpha c_1 f_{\mathcal{K},1} +\alpha^{2} c_1 c_2 f_{\mathcal{K},2}  & 1+\alpha^{2} c_1^{2} f_{\phi}  \\ \end{pmatrix}   
    \begin{pmatrix} \frac{1}{(J-2)r +3 r_{g}} \left[ -r^{2} \mathcal{K} + \frac{i \sqrt{A B} r }{\omega} H_1 \right]   \\ \addlinespace \hat{\phi} \end{pmatrix},
  \end{aligned}
\end{equation}
where $J=\ell (\ell+1)$ and
\begin{align}
    f_{h_1}^{-}&= \frac{ \left(4 x^3+3 x^2+3 x-7\right)}{4 x^6},\\
    f_{\mathcal{K}}^{+}&=\frac{1}{58320 x^6 \left[ (J-2) x+3 \right] ^{2}} \bigg[ 40 (J-2)^3 \left(2 J^4-19 J^3+66 J^2-46 J-133\right) x^7 \nonumber \\ & \quad +90 \left(4 J^6-54 J^5+300 J^4-772 J^3+630 J^2+615 J-902\right) x^6 \nonumber \\& \quad+30 \left(8 J^5-92 J^4+416 J^3-712 J^2-83 J+902\right) x^5 \nonumber \\ & \quad -45 \left(4 J^4-38 J^3+294 J^2-4439 J+7294\right) x^4 \nonumber \\ & \quad +27 \left(8 J^3-168 J^2+549 J+17060\right) x^3 -162 \left(2 J^2-98 J-163\right) x^{2} \nonumber \\ & \quad-486 (241 J-608) x -335340 \bigg]  \nonumber \\ & \quad-\frac{1}{4374} (J-2)^2 \left(2 J^4-19 J^3+66 J^2-46 J-133\right) \log \left[1+\frac{3}{(J-2) x}\right], \\
    f_{\phi,1} ^{+} &= -\frac{1}{x^2 \left[ (J-2) x+3 \right] } , \\
    f_{\phi,2}^{+} &=  - \frac{2+3 x + 6 x^{2}}{3 x^{5} \left[ \left( J-2 \right) x+3 \right] }, \\
    f_{\phi} &= -\frac{1}{29160 x^{6} \left[ \left( J-2 \right) x+3 \right] } \bigg[ 20 (J-2)^2 \left(2 J^4-19 J^3+66 J^2-46 J-133\right) x^6 \nonumber \\ & \quad+30 \left(2 J^5-23 J^4+104 J^3-178 J^2-41 J+266\right) x^5 \nonumber \\ & \quad -30 \left(2 J^4-19 J^3+66 J^2-46 J-295\right) x^4 \nonumber \\ & \quad +45 \left(50 J^3-231 J^2-135 J+1010\right) x^3 -243 \left(14 J^2-J+81\right) x^2 \nonumber \\ & \quad +486 (89 J-228) x + 46170 \bigg] + \frac{1 }{4374 x^{2} \left[ \left( J-2 \right) x+3 \right]} (J-2)^2\nonumber \\ & \quad  \left(2 J^2-5 J-7\right) \bigg[ \left(J^3-9 J^2+33 J-38\right) x^3  +3 \left(J^2-7 J+19\right) x^2 \nonumber \\ & \quad +54\bigg] \log\bigg[ 1+ \frac{3}{(J-2) x} \bigg], \\
    f_{\mathcal{K},1} &=  \frac{1}{54 x^{4}} \bigg[ 2 \left(2 J^3-9 J^2+3 J+14\right) x^3-3 \left(2 J^2-5 J+2\right) x^2 +3 (4 J-5) x  \nonumber \\ &\quad -108\bigg] -\frac{1}{81} (J-2)^2 \left(2 J^2-5 J-7\right) \log \bigg[ 1+\frac{3}{\left( J-2 \right) x} \bigg] , \\
    f_{\mathcal{K},2} &= \frac{1}{3061800 x^{7}} \bigg[ 140 \left(50 J^6-786 J^5+5823 J^4-21173 J^3+32253 J^2+9747 J \right. \nonumber \\ &\quad \left. -50338\right) x^6  -210 \left(50 J^5-686 J^4+4451 J^3-12271 J^2+7711 J+7430\right) x^5 \nonumber \\ & \quad  +210 \left(100 J^4-1172 J^3+6558 J^2-11426 J-1517\right) x^4 \nonumber \\ & \quad -945 \left(50 J^3-486 J^2+2307 J-1099\right) x^3   +1134 \left(100 J^2-772 J+8893\right) x^2 \nonumber \\ & \quad -11340 (25 J-371) x  +2259900 \bigg]  + \frac{1}{65610} (J-2)^2 \left(50 J^5-686 J^4 \right. \nonumber \\ & \quad \left.+4451 J^3-12271 J^2 +7711 J+25169\right) \log \bigg[ 1+\frac{3}{\left( J-2 \right)  x} \bigg] ,
\end{align}
we can determine the odd and even master equations up to $\mathcal{O}(\alpha^2)$, which are displayed in the form of Eqs.~\eqref{mastereqodd} and~\eqref{mastereqeven}. In Eqs.~\eqref{mastereqodd} and~\eqref{mastereqeven}, the GR potentials are
\begin{align}
    V^-_{\text{GR}}&= \frac{A_0}{r_g^2}\left(\frac{J}{x^2}-\frac{3}{x^3}\right)  ,\\ 
    V^+_{\text{GR}}&=\frac{A_0}{r_g^2}\frac{1}{x^3 [(J-2) x+3]^2}\bigg[(J-2)^2 J x^3 +3 (J-2)^2 x^2+9 (J-2) x+9\bigg]  ,\\ 
    V_{\text{GR}}^{(0)}&=\frac{A_0}{r_g^2} \left( \frac{J}{x^2}+\frac{1}{x^3} \right)  ,
\end{align}
where $A_0=1-1/x$. In Eq.~\eqref{mastereqodd}, the correction on the odd potential is 
\begin{align}
  V^-&=  \frac{1}{x^{5}} A_0  \left( 2 x^2+ 3 x+4\right) \omega ^2  + \frac{1}{3840 r_g^2 x^{10}}  \bigg[2880 x^6 -160 (71 J-567) x^5   \nonumber \\ &\quad +160 (73 J-774) x^4  +48 (11 J+745) x^3  +384 (61 J-1472) x^2  \nonumber \\ &\quad  -32 (685 J-32631) x   -491520 \bigg].
\end{align}
In Eq.~\eqref{mastereqeven}, the correction matrices for the even master equations are
\begin{align}
    \tilde{\mathbb{V}}_{1}=\begin{pmatrix} 0 &  c_1 V_{12 ,1} \\ \addlinespace c_1 V_{21 ,1} & c_2 V_{22 ,1}  \\\end{pmatrix} \qquad \text{and} \qquad \tilde{\mathbb{V}}_{2}=\begin{pmatrix} c_1 ^{2} V_{11 ,2} &  c_1 c_2 V_{12 ,2}  \\ \addlinespace c_1 c_2 V_{21 ,2}  & c_1 ^{2} V_{22 ,2}  \\\end{pmatrix},
\end{align}
where
\begin{align}
   V_{11 ,2}&= \frac{A_0}{60 x^5 [(J-2) x+3]^2} \bigg[ 30 (-2 + J) x^5 +40 \left(3 J^2-10 J+8\right) x^4 \nonumber \\ &\quad  +15 \left(12 J^2-3 J-34\right) x^3  +6 \left(40 J^2+4 J-3\right) x^2 +6 (215 J-196) x +1800 \bigg] \omega^{2} \nonumber \\ &\quad + \frac{1}{2160 r_{g}^{2} x^{10} [(J-2) x+3]^4} \bigg[ -20 (J-2)^3 \left(16 J^3-72 J^2+51 J+193\right) x^{10} \nonumber \\ &\quad+10 (J-2)^3 \left(32 J^3-975 J^2-813 J+6134\right) x^9 \nonumber \\ &\quad -30 (J-2)^2 \left(16 J^5-136 J^4+93 J^3+1008 J^2+7119 J-17278\right) x^8 \nonumber \\ &\quad +\left(480 J^7-10240 J^6+74337 J^5-203091 J^4+290038 J^3-1361064 J^2 \right. \nonumber \\ &\quad \left. +4237632 J-3886720\right) x^7 +\left(4240 J^6-44544 J^5+270042 J^4-908485 J^3 \right. \nonumber \\ &\quad \left. +2320044 J^2-5038140 J+4868200\right) x^6 \nonumber \\ &\quad +3 \left(290 J^5-32826 J^4+540715 J^3-2403814 J^2+4455660 J-3213578\right) x^5 \nonumber \\ &\quad -3 \left(1700 J^4+422486 J^3-3503670 J^2+8790359 J-7083286\right) x^4 \nonumber \\ &\quad+9 \left(39430 J^3-738504 J^2+2893344 J-3122597\right) x^3 \nonumber \\ &\quad  +54 \left(28330 J^2-225865 J+370936\right) x^2  +162 (13180 J-43223) x +874800 \bigg]  \nonumber \\ &\quad  + \frac{A_0}{81 r_{g}^{2} x^{5} [(J-2)x+3]^{2}} (J-2)^2 \left(2 J^2-5 J-7\right) \bigg[ 2 (J-2) x^4  + 3 x^3  \nonumber \\ &\quad+3 (J-2)^2 x^2 +13 (J-2) x+15\bigg], \\
   V_{12 ,1} &= \frac{A_0}{r_{g}^{2} x^{5} [(J-2)x+3]^{2}} \bigg[ 2 (J-2) x^4+3 (J-2)^2 x^2+13 (J-2) x+3 x^3+15 \bigg] ,\\
   V_{12 ,2} &= \frac{A_0}{30 r_{g}^{2} x^{9} [(J-2)x+3]^{2}} \bigg[ 146 (J-2) x^7+219 x^6+180 (J-2)^2 x^4  \nonumber \\ &\quad +90 \left(J^2+5 J-14\right) x^3+ 3 \left(20 J^2+58 J+119\right) x^2+(280 J-74) x+330 \bigg] , \\
   V_{21 ,1}&= \frac{2 A_0 (x+2)}{x^{3}}  \omega^{2}+\frac{A_0}{162 r_{g}^{2} x^{7} [(J-2)x+3]^{2}} \bigg[-6 (J-2)^2 \left(8 J^3-12 J^2-21 J \right. \nonumber \\ &\quad \left. +53\right) x^5  + 162 (J-2)^2 \left(3 J^2-7 J+1\right) x^4  \nonumber \\ &\quad +9 \left(356 J^3-1971 J^2+3684 J  -2332\right) x^3  +54 \left(227 J^2-1211 J+1532\right) x^2 \nonumber \\ &\quad +729 (52 J-147) x+46656 \bigg] + \frac{2 (2 J-7) A_0}{81 r_{g}^{2} x^{2} [(J-2)x+3]^{2}}   \left(-J^2+J+2\right)^2 \nonumber \\ &\quad  \bigg[ 2 (J-2) x+3 \bigg] \log \bigg[ 1+\frac{3}{(J-2)x} \bigg] , \\
   V_{21 ,2} &=  \frac{A_0}{15 x^{7}} \left( 73 x^4+146 x^3+219 x^2+112 x+50 \right) \omega^{2} \nonumber \\ &\quad  +\frac{A_0}{1530900 r_{g}^{2} x^{10} [(J-2)x+3]^{2}} \bigg[     70 (J-2)^2 \left(200 J^6-2544 J^5+15060 J^4 \right. \nonumber \\ &\quad \left.-31280 J^3-18240 J^2+78303 J  -58975\right) x^8  -3725190 (J-2)^2 (J-1) x^7  \nonumber \\ &\quad +105 \left(100 J^6-1272 J^5+95010 J^4 -575998 J^3  +1188465 J^2-657732 J \right. \nonumber \\ &\quad \left. -316268\right) x^6 \nonumber \\ &\quad  +1260 \left(515 J^5-677 J^4  +37547 J^3 -243997 J^2  +517714 J-347725\right) x^5  \nonumber \\ &\quad +3402 \left(1825 J^4-2438 J^3+34695 J^2-300521 J+462421\right) x^4 \nonumber \\ &\quad +1134 \left(20050 J^3-15606 J^2+252753 J-1349411\right) x^3\nonumber \\ &\quad  +1215 \left(59090 J^2-67616 J +335021\right) x^2 + 131220 (1895 J-2291) x \nonumber \\ &\quad+332970750 \bigg]   +\frac{A_0}{32805 r_{g}^{2}x^{6} [(J-2)x+3]^{2}} (J-2)^2 (J+1) \bigg[ \nonumber \\ &\quad 2 \left(50 J^6-786 J^5+5823 J^4-21173 J^3+32253 J^2+9747 J-50338\right) x^5 \nonumber \\ &\quad +3 \left(50 J^5-686 J^4+4451 J^3-12271 J^2+7711 J+25169\right) x^4 \nonumber \\ &\quad  +2430 (J-2)^2 (2 J-7) x^2  + 14580 \left(2 J^2-11 J+14\right) x \nonumber \\ &\quad +21870 (2 J-7) \bigg]  \log \bigg[ 1+\frac{3}{(J-2)x} \bigg] ,\\
   V_{22 ,1}&= -\frac{6 A_0}{r_{g}^{2} x^{6}},\\
   V_{22 ,2}&= -\frac{1}{2160 r_{g}^2 x ^{10} [(J-2)x+3]^{2}} \bigg[ -20 \left(4 J^2-13 J+10\right)^2 x^8 \nonumber \\ &\quad +10 \left(32 J^4-217 J^3+345 J^2 +308 J-772\right) x^7 \nonumber \\ &\quad -30 \left(16 J^5-136 J^4+415 J^3-504 J^2+222 J-138\right) x^6 \nonumber \\ &\quad +\left(480 J^5 -5440 J^4+19823 J^3-18447 J^2-29938 J+44870\right) x^5 \nonumber \\ &\quad +2 \left(680 J^4-14608 J^3 +94539 J^2 -218932 J+172712\right) x^4 \nonumber \\ &\quad +3 \left(6870 J^3-138430 J^2+581797 J -671341\right) x^3 \nonumber \\ &\quad  +12 \left(18925 J^2-175009 J+316771\right) x^2  +18 (45385 J-173657) x+952560\bigg]  \nonumber \\ &\quad  -\frac{A_0}{81 r_{g}^{2} x^{5} [(J-2)x +3]^{2}} (J-2)^2 \left(2 J^2-5 J-7\right) \bigg[   2 (J-2) x^4+3 x^3\nonumber \\ &\quad+3 (J-2)^2 x^2  +13 (J-2) x+15 \bigg] .
\end{align}

For the coupled even modes, in order to use the WKB approximation to find the phase shifts, we need to determine the eigenvalues of the matrix $\omega ^{2} \mathbbm{1}- \tilde{\mathbb{V}}$ in Eq.~\eqref{mastereqeven}. We denote these two eigenvalues as $\omega^{2}-V^{+}$ and $\omega^{2}- V^{(0)}$. Up to $\mathcal{O}(\alpha^{2})$, we have $V^{+}=V^{+}_{\mathrm{GR}}+\alpha ^{2} c_1^{2}  \delta V^{+}_{1}$ and $V^{(0)}=V_{\mathrm{GR}}^{(0)}+ \alpha c_2 \delta V_2^{(0)}+\alpha^{2} c_1^{2} \delta V_1^{(0)}$, where
\begin{align}
    \delta V^+_{1}&= \frac{A_0}{60 (J+1) x^6 [2 (J-2) x+3] [(J-2) x+3]^{2}  } \bigg[ 60 (J-5) (J-2)^2 x^7 \nonumber \\ &\quad -10 \left(24 J^4-128 J^3 +159 J^2+159 J-314\right) x^6 \nonumber \\ &\quad -30 \left(6 J^4+45 J^3  -287 J^2+392 J  -92\right) x^5 \nonumber \\ &\quad -3 \left(40 J^4+376 J^3+627 J^2-5615 J+4894\right) x^4  \nonumber \\ &\quad +30 \left(14 J^3-366 J^2  +621 J-79\right) x^3 + 18 \left(385 J^2-2329 J+2416\right) x^2 \nonumber \\ &\quad +1620 (12 J-29) x+16200\bigg] \omega^{2} \nonumber \\ &\quad +\frac{1}{240 (J+1) r_{g}^{2} x^{11} [2(J-2)x+3][(J-2)x+3]^{4}} \bigg[ \nonumber \\ &\quad 120 (J-2)^4 \left(J^2-2 J-15\right) x^{12} \nonumber \\ &\quad -20 (J-2)^3 \left(107 J^4-138 J^3-474 J^2-250 J+1923\right) x^{11} \nonumber \\ &\quad +10 (J-2)^3 \left(254 J^4-2243 J^3-906 J^2+269 J+18118\right) x^{10} \nonumber \\ &\quad -2 (J-2)^2 \left(540 J^6-5433 J^5+10265 J^4+19682 J^3-26187 J^2 \right. \nonumber \\ &\quad \left. +242969 J -554410\right) x^9  \nonumber \\ &\quad +3 \left(360 J^8 -10304 J^7 +106849 J^6-519180 J^5+1317493 J^4 \right. \nonumber \\ &\quad \left. -1549958 J^3-563376 J^2+3943968 J -3312480\right) x^8 \nonumber \\ &\quad +\left(15620 J^7-365652 J^6 +3188400 J^5-13264679 J^4+28079109 J^3 \right. \nonumber \\ &\quad \left. -26074812 J^2   -1076848 J  +13399344\right) x^7 \nonumber \\ &\quad +3 \left(45270 J^6-878840 J^5+6117191 J^4-19798739 J^3 +30852172 J^2 \right. \nonumber \\ &\quad \left.-19218610 J +654778\right) x^6  \nonumber \\ &\quad   +9 \left(88600 J^5-1332986 J^4+7046280 J^3-16596395 J^2 +17108665 J \right. \nonumber \\ &\quad \left.  -5600874\right) x^5  \nonumber \\ &\quad +9 \left(340210 J^4-3831458 J^3 +14643168 J^2-22732079 J+11951965\right) x^4 \nonumber \\ &\quad +162 \left(47070 J^3-381259 J^2 +956143 J-747808\right) x^3 \nonumber \\ &\quad +486 \left(24880 J^2-131577 J+164663\right) x^2 +4860 (2312 J-6049) x \nonumber \\ &\quad +4665600\bigg], \\
    \delta V_2^{(0)}&=-\frac{6 A_0}{r_{g}^{2} x ^{6}},\\
    \delta V_1^{(0)}&= \frac{A_0 (x+2)}{(J+1)x^{6}[(J-2)x+3]} \bigg[ 2 (J-2) x^4+3 x^3+3 (J-2)^2 x^2+13 (J-2) x \nonumber \\ &\quad +15 \bigg] \omega^{2}  +\frac{1}{240 (J+1)r_{g}^{2}x^{11}[(J-2)x+3]} \bigg[ 120 \left(J^2-5 J+6\right) x^8  \nonumber \\ &\quad  +20 \left(37 J^3-142 J^2+148 J-33\right) x^7 \nonumber \\ &\quad  -10 \left(82 J^3-413 J^2+1064 J-1177\right) x^6 \nonumber \\ &\quad +6 \left(180 J^4-1069 J^3+1939 J^2+1248 J-6170\right) x^5 \nonumber \\ &\quad + \left(-1080 J^4+16608 J^3-72417 J^2+80496 J+34521\right) x^4 \nonumber \\ &\quad -4 \left(2455 J^3-28001 J^2+61403 J  -13711\right) x^3  \nonumber \\ &\quad   -2 \left(26095 J^2-129754 J+92191\right) x^2-180 (520 J-987) x-57600\bigg] .
\end{align}

For the odd mode, the leading-order sGB correction on its radial sound speed $(c_{s}^{-})^{2}$, defined as $ \alpha^{2} \Delta c_s^- \equiv (c_{s}^{-})^{2}-1$, is given by the coefficient of $\omega^{2}$ in $V^-$,
\begin{align}
    \Delta c_s^{-} = & \frac{c_1^2}{x^{5}} A_0  \left( 2 x^2+ 3 x+4\right).
\end{align}
For the coupled even modes, there are two characteristic sound speeds $(c_{s}^{+})^{2}$ and $(c_{s}^{(0)})^{2}$. These characteristic sound speeds are determined by the eigenvalues of the inverse matrix of the coefficient matrix of $\omega^{2}$ in $\omega ^{2} \mathbbm{1}-\tilde{\mathbb{V}}$. Only $(c_{s}^{+})^{2}$ will receive a leading-order correction. We denote the leading-order sGB correction on $(c_{s}^{+})^{2}$ as $ \alpha^{2} \Delta c_s^+ \equiv (c_{s}^{+})^{2}-1$, where
\begin{align}
    \Delta c_s^{+} &=  \frac{c_1^2}{60 x^{5}[(J-2)x+3]} A_0   \bigg[30 (J-2) x^5   + 40 \left(3 J^2-10 J+8\right) x^4  \nonumber \\ &\quad  +15 \left(12 J^2-3 J-34\right) x^3+6 \left(40 J^2+4 J-3\right) x^2+6 (215 J-196) x  +1800 \bigg] .
\end{align}

\bibliography{biblio}

\end{document}